\documentclass[runningheads]{llncs}
\usepackage{amssymb,amsmath}
\usepackage{pstricks}
\usepackage{pst-node}
\usepackage{epsfig}
\usepackage{latexsym}
\begin{document}
\pagestyle{headings}

\def\imply{\Rightarrow}     

\date{\ }

\title{Paraconsistency of Interactive Computation
\thanks{Originally published in proc. PCL 2002, a FLoC workshop;
eds. Hendrik Decker, Dina Goldin, J{\o}rgen Villadsen, Toshiharu Waragai
({\tt http://floc02.diku.dk/PCL/}).}
}

\author{Dina Goldin\inst{1}\fnmsep\thanks{Supported by NSF Grant \#IRI-0296195 and
by a grant from the Connecticut Research Foundation.}    
\and Peter Wegner\inst{2}}

\institute{U. Conn, Storrs, CT \\ \email{dqg@cse.uconn.edu}
\and Brown University, Providence, RI \\ \email{pw@cs.brown.edu}}

\maketitle

\begin{abstract}
The goal of computational logic is to allow us to 
{\em model} computation as well as to {\em reason} about it.
We argue that a computational logic must be able to model
{\em interactive computation}. 
We show that first-order logic
cannot model interactive
computation due to the {\em incompleteness} of interaction.
We show that interactive computation is 
necessarily {\em paraconsistent}, able to model both a fact
and its negation, due to the role of the world
(environment) in determining the course of the computation. 
We conclude that paraconsistency is a necessary property for a logic
that can model interactive computation.
\end{abstract}

\section{Introduction}

The
paradigm shift from algorithms to interactive computation captures the
technology shift from mainframes to workstations and now to wireless devices
and intelligent appliances,
from number-crunching to embedded systems and graphical user interfaces,
and from procedure-oriented to object-based and distributed computation.
{\em Interactive computation}, modeled by
interactive systems, is a more powerful paradigm 
of problem solving than {\em algorithmic computation}, modeled by
Turing Machines.

The view that TMs completely express the intuitive notion of
computing is a common misinterpretation of Church-Turing Thesis.
This view has led computer science theorists
to assume that questions of expressiveness
of finite computing agents had been settled once and for all,
so that exploration of alternative models of computability has seemed
unnecessary.
The claim that interactive computing agents are more {\em expressive} than
Turing Machines, meaning {\em able to solve a larger set of
computational problems},
requires a reexamination of fundamental assumptions about models
of computation, about computation itself, and about computational
logics.  

In section~\ref{sec:power}, we introduce {\em interactive computation}
and demonstrate that it is more {\em powerful} than
algorithmic computation of Turing Machines,
in that it can solve a wider class of computational problems.
In section~\ref{sec:logic}, we show that classical first-order logic
cannot model interactive computation.
In section~\ref{sec:consistency}, 
we examine areas of computing where the principle of {\em consistency}
plays a major role;
in each case, these areas are fundamentally not interactive.
Finally, in section~\ref{sec:extending},
we consider embedding logic models in the broader context 
of interactive models, and conclude that 
paraconsistency is necessary to model interaction.

\section{The power of interactive computation}
\label{sec:power}

In this section, we introduce {\em interactive computation}
and demonstrate that it is more {\em powerful} than
the algorithmic computation of the Turing Machines,
in that it can solve a wider class of computational problems.

\subsection{Interactive computation}

The notion of {\em algorithms} well precedes computer science, 
having been a concern of mathematicians for centuries.  It
can be dated back to a 9th century treatise by a Muslim 
mathematician Al-Koarizmi, after whom algorithms were named.
\begin{quote}
{\bf Algorithm:} systematic procedure that produces --
in a finite number of steps -- the answer to a question 
or the solution of a problem~\cite{Brit}.
\end{quote}
{\em Algorithmic computation}, modeled by Turing Machines, was adopted
as the traditional model of computation when computer science
emerged as a new academic discipline in the 1960s.     
\begin{quote}
{\bf Algorithmic computation:}
the computation is performed in a {\em closed-box} fashion, transforming
a finite input, determined by the start of the computation, to a finite
output, available at the end of the computation, in a finite
amount of time.
\end{quote}
Whereas algorithmic computation is like the {\em data processing mode}
associated with punch cards of the 1950s,
today's computational systems are {\em interactive}.
In these systems,
which are expected to remain up and running continuously,
input is consumed and output is generated throughout the computation.
More importantly, there is an interdependence of input and output:
output can influence later input, and vice versa. 

\subsection{The \emph{WH} problem}
\label{sec:WH}

Let us consider an {\em automatic car}  
whose task is to drive us across town 
from point \emph{W} (work) to point \emph{H} (home);
we shall refer to it as the {\em WH problem}.  
The output for this problem should be 
a time-series plot of signals to the car's
controls that enable it to perform this task autonomously.
At issue is what form the inputs should take.

The car can be equipped with a map of the city. 
In the {\em algorithmic} scenario, where all inputs 
are provided {\em a priori} of computation, 
this map needs to include every pothole
and even grain of sand along the road. By the principles
of {\em chaotic behavior}, such elements can greatly affect the
car's eventual course -- like the
Japanese butterfly that causes a tsunami on the other end of the world.

In a {\em static} world, such a map is in principle obtainable;
but ours is a {\em dynamic} environment. The presence of
mutable physical elements such as the weather affect the 
car's course both directly (e.g., the wind blowing at the car)
and indirectly
(e.g. sand shifting location in the path of the car as a result of
blowing wind).  
It is arguable whether the elements of the weather can be precomputed
to an accuracy required to produce the proper answer to the 
\emph{WH} problem.

We can remain optimistic until we remember that the world also 
includes humans, as pedestrians or drivers.  To avoid collisions,
we must precompute the motion of everyone who
might be near the car as it drives. 
An assumption that human actions can be computed ahead of time
is tantamount to an assertion of {\em fatalism} -- a doctrine 
that events are fixed in advance so that human beings are 
powerless to change them -- clearly
beyond the purview of any computer scientist.  Therefore,
we must assume it cannot be done.
\begin{quote}
{\it
The claim that computational tasks situated 
in a dynamic world that includes human agents are solvable algorithmically
is tantamount to a assertion of fatalism.
}
\end{quote}

\subsection{The power of interaction}

Nevertheless, the \emph{WH} problem
is solvable -- interactively.  In this scenario,
the inputs, or {\em percepts}~\cite{AIMA}, consist of a stream of
images produced by a video camera mounted on the car, as it is
driving from \emph{W} to \emph{H}. The signals to the car's controls are 
generated {\em on-line}
in response to these images, to avoid steering off the road
or running into obstacles.  This change in the scenario is akin
to taking the blindfolds off the car's electronic driver, who was
driving from memory and bound to a precomputed sequence of actions.
Now, he is aware of his environment and uses it to guide his
steering.

Note that a change of approach from algorithmic to interactive such 
as above involves much more than a restructuring of inputs.  
The {\em nature of computation} itself is
at stake. Algorithmic computation is {\em off-line}; 
it takes place {\em before} the driving begins.  Interactive computation
is {\em on-line}; it takes place {\em as} the car drives.
The values of inputs and outputs for interactive computation 
are interdependent; decoupling them,
such as by replacing the videocamera with a prerecorded videotape 
of the road, will usually lead to an incorrect solution.

This interpretation of the concept of computation is different from
the one that Church and Turing had in mind for their
famous thesis, where they specifically referred to computation of functions
over finite strings that encode natural numbers (i.e., algorithmic computation).
Interactive computation falls outside the bounds of the 
Church Turing thesis. The \emph{WH} example shows that interaction 
pulls us out of the
{\em Turing tarpit}, by providing solutions to problems that are
not solvable in the algorithmic setting.

\begin{quote}{\it
By allowing us to solve computational tasks that cannot
be solved algorithmically, interaction is shown to be a
more powerful computational paradigm.
}
\end{quote}

\section{Logic and interactive computation}
\label{sec:logic}

In this section, we discuss why classical first-order logic
cannot model interactive computation.
\subsection{Tradeoffs between reasoning and modeling}
\label{sec:constr}
\label{sec:RE}

Logic comprises both syntactic rules for noninteractively 
(in a closed-box fashion) proving {\em theorems} from {\em axioms} 
by {\em rules of
inference} (proof theory) and semantic notions of soundness and
completeness that relate syntactic processes of proof to semantic
properties of a modeled domain in which formulae are interpreted (model 
theory). A logic
is {\em sound} if all its theorems are true, and {\em complete} if all true
assertions of the modeled domain are theorems.

In sound and complete formal systems
such as first-order logic, every true assertion about the domain
has a finite proof that is constructed by applying the rules to the
axioms. With the number of rules and axioms also finite, 
the resulting set of assertions must be {\em recursively enumerable} (RE).  

\begin{quote}
{\bf Necessary condition for completeness}: A sound and
complete first-order logic (SCL)
can model only domains with a countable set of properties.  
\end{quote}

Completeness is a desirable mathematical property that certifies
formalizability of the modeled domain but
by the same token restricts expressiveness. 
We may wish to model classes of systems 
(mathematical, computational, or physical) which have a non-RE
set of properties.
Modeling is {\em observational} rather than {\em constructive},
based on {\em greatest fixpoints} rather than {\em least fixpoints}
~\cite{coind}. 
The gap between least- and greatest-fixpoint semantics is 
also the gap between operational (algorithm) and
denotational (observation) semantics.  It is also the same
as the gap between {\em deduction} and {\em abduction}.

Greatest fixpoints allow us to define larger domains.
Lacking a constructive foundation, we cannot systematically
enumerate the members of such domains.  However, it can be observed whenever
two members are distinct;  distinguishability certificates
are always finite.  The {\em bisimulation relation} for 
labeled transition systems is an example of a relation
with greatest fixpoint semantics and finite distinguishability
certificates. 

\begin{quote}
{\bf Sufficient condition for incompleteness:} Any system
with an uncountable set of properties cannot be expressed by a
sound and complete logic.
\end{quote}

As a result, there are tradeoffs between reasoning and modeling: 
purely rule-based syntactic reasoning is
too weak to capture all semantic properties of mathematical, computational, 
or physical systems. Model
builders must choose between complete formalizability and modeling
power. Turing Machines focus on formalizability
at the expense of modeling power.

(As an aside, this focus on formalizability is intentional; TMs
were invented as part of Turing's proof
that Hilbert's challenge to formalize mathematics cannot be 
carried out.  The {\em halting problem} presented a constructive
counterexample to this challenge~\cite{Tur36}.  Turing himself
never claimed that TMs formalize all of computation, and in fact
made claims that suggested the opposite~\cite{new-cacm}.)

\subsection{Incompleteness of interactive systems}

Godel showed that
arithmetic over the integers cannot be completely formalized by logic,
because not all its properties are syntactically expressible as
theorems. We
extend Godel's reasoning to show that interactive and empirical systems 
are likewise incomplete because
they have too many properties to be expressible as theorems of a sound
and complete logic.

As discussed above, sound and complete first-order logics (SCLs) have an RE
set of theorems and can formalize only semantic domains with
a countable number of distinct properties.
Incompleteness occurs when the number of true facts or observable
properties of a system cannot be recursively enumerated by theorems. 
For example, arithmetic over the integers is 
incomplete because its true properties
cannot be recursively enumerated; that is, for any sound formalization we can
find a true property of the integers that cannot be proved. 

For interactive systems or {\em interaction machines} (IMs), properties 
correspond to {\em observations}, and
the number of properties corresponds to the number of distinct
observations. 
Algorithms and TMs have
only a countable set of properties, while IMs have an
open-ended set of properties that
is generally not countable and therefore cannot be expressed as theorems of a
complete first-order logic.
Godel's incompleteness result holds for any
domain whose true properties cannot
be recursively enumerated, including empirical domains modeled by
IMs.

Expressiveness for IMs is defined in terms of the ability of
observers to make observational distinctions~\cite{TCS}. 
Observational expressiveness
allows behavior of algorithms and IMs to be measured by 
the same metric. 
The class of systems or programs which admits finer observational 
distinctions is the one capable of a greater range of behaviors,
and hence able to express solutions to a larger set of problems.
From this viewpoint, 
TM behavior is completely characterized by a single observation, while IM
behavior may depend on unbounded sequences or patterns of observations. 
This gives another proof of greater expressive power of interaction.

Completeness restricts the expressiveness of domains modeled by an SCL: 
domains whose 
properties are not countable are necessarily incomplete. 
IMs, based on infinite streams of inputs and outputs
created in partnership with an (uncomputable) environment,
have an uncountable set of computations: the distinct streams
of an IM are not enumerable.

By viewing Godel's incompleteness result as a corollary of the more
general result that SCLs cannot
model systems with an uncountable set of properties we gain insights 
into the nature of incompleteness.
Incompleteness is seen to be a ubiquitous phenomenon that applies not
only to mathematical systems like
the integers but also to IMs.

\begin{quote}
{\em The set of behaviors of an interactive system cannot in general be
formalized by any sound and complete logic.}
\end{quote}

Incompleteness shows the limitations of syntactic formalisms in
expressing semantic behavior. The
completeness/incompleteness dichotomy distinguishes algorithms from
IMs, {\em closed} from
{\em open} systems, and rationalist from empiricist models~\cite{monist}. 
Formal incompleteness of computing systems 
is related to descriptive incompleteness of physical models of the 
real world.

Interactive problem solving (section~\ref{sec:WH})
is a first-class form of computation that 
should be included in any intuitive
notion of computability.  Hence, logics need to 
go beyond soundness and completeness to capture computation.

\subsection{Logic for physical theories}

The relatively recent development of {\em paraconsistent logic} 
challenges the logical principle that contradictory premises
lead to an explosion of meaningless conclusions.
\begin{quote}
{\bf Paraconsistency:} 
Let $\imply$ be a relation of logical consequence.
Let us say that $\imply$ is {\em explosive} iff for every formula
$A$ and $B$, $\{A, \lnot A\} \imply B$.  Classical logic and most other
standard logics are explosive.  A logic is said to be {\em paraconsistent}
iff its relation of logical consequence is not explosive
~\cite{Priest}.
\end{quote}
The most telling reason for paraconsistent logic 
is the fact that there are non-trivial theories which are 
inconsistent. Some of the best examples of such theories come from the 
history of physics.

The interactive view of computing allows us to bridge the gap
between closed-box algorithmic computing and real physical
systems. Interactive computation spans the range from closed-box
algorithmic computation at one end to situated environment-aware
computing agents (e.g. the car in the \emph{WH} problem) in the middle, 
to the behavior of real world objects at the other extremum.  

The notion of {\em observation} as a metric of expressiveness
and equivalence for IMs is a natural extension
of the role of observation in the formulation of physical
theories~\cite{monist}.  The
idea that physical objects are not completely describable or knowable but
that they may have describable parts or views is a basic tenet of the
scientific method.  An analogous idea for IMs
is that {\em there is no silver bullet} -- 
when testing a software system, one can never be
sure of having found all the bugs.
\begin{quote}
{\em Just as physical theories provide an important 
motivation for paraconsistent logics, so does the theory
of interactive computation.}
\end{quote}

\section{Consistency in computation}
\label{sec:consistency}

In this section, we examine some of the areas of computing which
have been founded on the principle of {\em consistency}, i.e., 
that it is impossible for $A$ and $\lnot A$ to coexist.
We show that in each case, these areas are fundamentally
not interactive.

\subsection{Logic programming}

Logic programming corresponds to 
closed-system computing, whereas interactive systems
are inherently open. The Japanese 5th generation computing project 
that aimed to reduce computation to logic programming proved a failure.
It was shown in~\cite{pw93} that this was due not to lack of
cleverness on the part of logic programming researchers, but to
the theoretical impossibility of such a reduction. 

The key argument is the inherent incompatibility between
reactiveness of interactive computation, realized by committed choice 
at each interaction step, and logical completeness, realized by
backtracking. Committed choice is inherently incomplete because 
commitment cuts off branches of the proof tree
that might contain the solution.
However, commitment is essential to interactive computation, 
since execution of an observable action by a system cannot be reversed.
The power of interaction
can be achieved only by sacrificing logical completeness. 
Pure logic programming is inherently too weak to serve 
as a model for interactive computation.

\subsection{Data Management}

Despite significant overlaps, 
the research areas of {\em databases} (DB) and {\em information systems} (IS)
are based on fundamentally different assumptions, due to the 
different jobs they perform.  The ``{\em job
of a database}'' is to store data and answer queries. This entails
addressing issues like data models, schema design, handling
distributed data, maintaining data consistency, query evaluation, etc.

Given a query and a database instance, the {\em behavior} of a
database system consists of answering queries, which is {\em algorithmic} in
nature. In particular, a database management system
implements an algorithmic mapping of the form:
\begin{quote}
\it{(DB contents, user query)} $\rightarrow$ \it{(DB contents', query
answer)}.
\end{quote}
On the other hand, the {\em job of an IS} is to
provide information-based services,
which entails considerations that span the
life cycle of the larger system. Services over time are {\em interactive} in
nature, involving user sessions with one or more users. 

It is not surprising then that consistency is taken for granted
in database research, given its algorithmic nature. 
However, most large contemporary database systems are to some degree
IS's, providing additional services beyond simply
storing data and running queries and updates~\cite{db-is}.  When lines 
between the two are blurred in this fashion, the tension between
consistency and interaction can result.

\subsection{Artificial Intelligence}

Classical logic has always played a prominent role in {\em Artificial 
Intelligence} (AI), where a typical problem would be to form a plan
or to solve a problem, given a complete description of the 
``world'' in question.  Closed-box computation was the {\em modus operandi},
even in problems like rearranging blocks
in {\em Shakey's world}~\cite{AIMA}
that were meant to offer a simplified version of situated
real-world behavior.

Many researchers came to realize that algorithmic approaches
cannot scale up to solutions for tasks involving the 
real world~\cite{Brooks}, like the \emph{WH} problem; a paradigm
shift was needed.
The new agent-based approach to AI~\cite{AIMA} is interactive.
The change in approach has reinvigorated this field, breaking
the inertia that gripped it in the '80s, and restoring our
optimism in the promises that AI was making when originally
founded.

Not surprisingly, classical logic does not play the same
prominent role in the new AI.  The emphasis has shifted
to probablistic approaches, with such models as 
{\em neural networks}, {\em hidden Markov models} and 
{\em belief networks}~\cite{AIMA}.  We expect that 
paraconsistent logics can also play a significant role in
the new AI.

\section{Extending logic to interaction}
\label{sec:extending}

By embedding logic models in the broader context of interactive models, 
we gain insight into the limits
of logic in modeling autonomous external worlds.  
We also understand why paraconsistency is necessary to model interaction.

The number of interpretations of function and predicate symbols of
first-order logic ({\em fol\/}) is nonenumerable.  However, the set of theorems
for a fixed interpretation is recursively enumerable (section~\ref{sec:RE}).
By assuming a static interpretation of the world that is fixed
prior to the start of the computation, algorithmic models 
have a recursively enumerable set of behaviors.
By contrast, the nonenumerability of interpretations of {\em fol}s
reflects the nonenumerable external worlds (environments) of 
interactive computation, which are neither static nor known {\em a priori}.
Interactive models replace complete semantics of {\em fol}s by 
weaker abstractions that sacrifice completeness for expressiveness.
\begin{quote}
{\em Conjecture:} Late (dynamic) binding of interpretations in logic is a 
form of interactive modeling.
\end{quote}

Interaction extends models so that the interpretation of nonlogical 
symbols evolves during inference to
take account of new data. However, interactive discovery of facts 
negates the fundamental property that
true facts always remain true. {\em Nonmonotonic logics} permit 
reasoning about incompletely described modeled worlds. 
As the system gains information that contradicts earlier 
inferences (beliefs), it may retract these inferences.

The contrast between the constructive paradigm of algorithmic
models and the observational paradigm 
of interaction 
is the contrast between {\em deduction} and {\em abduction}
(section~\ref{sec:constr}).
Abduction is open-ended and subject to revision; abductive 
reasoning is {\em paraconsistent} in that it is possible for incompatible
inferences to be drawn from the same premises.

Abduction and nonmonotonicity are not only desirable, but required
for interactive computation, where input, inference, and action
are interleaved.  
Paraconsistency, the embodying of contradictory information,
is necessary for abductive and nonmonotonic reasoning, and hence
for interaction.

\section{Conclusion}

The goal of computational logic is to allow us to
{\em model} computation as well as to {\em reason} about it.
A useful computational logic must be able to model
{\em interactive computation}.
We showed that first-order logic
cannot model interactive
computation due to its {\em incompleteness}.
Interactive computation is
necessarily {\em paraconsistent}, able to model both a fact
and its negation, due to the role of the world
(environment) in determining the course of the computation.
We noted that interactive computation shares much in common
with scientific theories, providing a common motivation for
the development of paraconsistent logic.
We concluded that paraconsistency is a necessary property for a logic
that can model interactive computation.

\bigskip

\section*{Acknowledgements}

We thank David Keil for his careful reading of earlier versions
of this work.


\bigskip

\begin{thebibliography}{10}    

\bibitem{Brooks}
Rodney A. Brooks.
\newblock {\em Intelligence Without Reason}.
\newblock MIT A.I. Memo, 1991.
\newblock {\tt http://www.ai.mit.edu/people/brooks/}

\bibitem{Brit}
{\em Encyclopaedia Britannica}.

\bibitem{db-is}
Dina Goldin, Srinath Srinivasa, Bernhard Thalheim. 
\newblock {\em Information Systems = Databases + Interaction: Towards Principles of Information System Design}.
\newblock ER2000, Salt Lake City, October 2000. 

\bibitem{Priest}
Graham Priest, Koji Tanaka.
\newblock Paraconsistent Logic.
\newblock {\em Stanford Encyclopedia of Philosophy}.
\newblock 2000.
\newblock {\tt http://plato.stanford.edu/}

\bibitem{AIMA} 
Stuart Russell, Peter Norvig. 
\newblock {\em Artificial Intelligence: A Modern Approach}. 
\newblock Addison-Wesley, 1995.

\bibitem{Tur36}
Alan Turing.
\newblock On Computable Numbers with an Application to the Entscheidungsproblem.
\newblock {\em Proceedings of the London Mathematical Society}, 2(42), pp. 173-198, 1936.

\bibitem{pw93}
Peter Wegner.
\newblock Tradeoffs Between Reasoning and Modeling.
\newblock In {\em Research Directions in Concurrent Object-Oriented Computing}, Agha, Wegner, Yonezawa (Eds.).
\newblock MIT Press, 1993.

\bibitem{TCS}
Peter Wegner.
\newblock Interactive Foundations of Computing.
\newblock {\em Theoretical Computer Science}, February 1998.

\bibitem{monist}
Peter Wegner.
\newblock Towards Empirical Computer Science.
\newblock {\em The Monist}, Issue on the Philosophy of Computation, January 1999.

\bibitem{coind}
Peter Wegner and Dina Goldin.
\newblock Coinductive Models of Finite Computing Agents.
\newblock Proc. Coalgebra Workshop (CMCS '99).
\newblock {\em Electronic Notes in Theoretical Computer Science}, Volume 19, March 1999.

\bibitem{new-cacm}
Peter Wegner and Dina Goldin.
\newblock {\em Computation Beyond Turing Machines.}
\newblock 2002.
\newblock Accepted for publication.

\end{thebibliography}
\end{document}